\documentclass[journal]{IEEEtran}

\usepackage{graphicx}
\usepackage{cite}
\usepackage{amsmath,amssymb,amsfonts}
\usepackage{tabularx}
\usepackage{booktabs}
\usepackage{multirow}
\usepackage{array}
\usepackage{url}
\newtheorem{definition}{Definition}

\usepackage{capt-of}

\begin{document}

\title{Data Center Load Accommodation under QoS and Checkpoint-Induced Frequency Security Constraints}

\author{Ruolan Cui,~\IEEEmembership{Student Member,~IEEE,}
Ziang Liu,~\IEEEmembership{Student Member,~IEEE,}
Chenxi Hu,~\IEEEmembership{Member,~IEEE,}\\
Francis Yunhe Hou,~\IEEEmembership{Fellow,~IEEE}

\thanks{
This work was supported in part by the National Key R\&D Program of China under Grant 2023YFA1011301, and in part by RGC/GRF under Grant 17201524.

Ruolan Cui, Ziang Liu and Francis Yunhe Hou are with the Department of Electrical and Computer Engineering, The University of Hong Kong, Hong Kong.
Chenxi Hu is with the Department of Computer Science, The Hang Seng University of Hong Kong, Hong Kong.
}
}

\maketitle

\begin{abstract}
The rapid growth of artificial intelligence workloads is accelerating the expansion and grid interconnection of data centers (DCs). Existing data center load accommodation (DCLA) assessments primarily focus on grid and resource adequacy and often imply that greater workload flexibility always increases accommodable capacity. However, workload flexibility has a dual effect. Workload deferral and temporal shifting can alleviate network congestion, whereas synchronized checkpointing of training workloads can cause large and rapid load reductions, threatening frequency stability. Therefore, DCLA cannot be determined solely by grid adequacy. This paper develops an optimal DCLA method that jointly accounts for workload quality of service (QoS), checkpoint-induced frequency security (FS), and grid adequacy. First, a QoS-aware model characterizes workload-dependent flexibility while ensuring service feasibility. Second, checkpoint-induced FS constraints are developed for periodic and event-triggered power variations, with key frequency-security limits reformulated in algebraic form. Third, a two-stage lexicographic model is developed to maximize DCLA and minimizes the required energy storage (ES) deployment, and then optimizes system operation under the given DC and ES siting and sizing decisions. Case studies show that the proposed method avoids a 40.3\% overestimation of DCLA compared with the grid-adequacy benchmark. QoS is the primary limiting factor, while checkpoint-induced frequency-security constraints become more restrictive as training workloads increase. These findings highlight that reliable DCLA should account for both its grid-support value and the associated QoS and FS risks.

\end{abstract}


\IEEEpeerreviewmaketitle

\section{Introduction}

\IEEEPARstart{T}{he} rapid growth of artificial intelligence (AI) and large language model training is driving a substantial increase in data center (DC) electricity demand, making DCs one of the fastest-growing and most spatially concentrated emerging loads in power systems~\cite{nerc_large_loads2025,vercellino2026_ai_power}. Large-scale DC interconnection places increasing pressure on transmission corridors, substation capacity, resource adequacy, and interconnection processes~\cite{epri_dcflex2025}, while grid expansion often lags behind load growth because of siting, permitting, investment approval, and long construction lead times~\cite{nerc_gaps_2026}. Therefore, determining how much DC load can be securely connected to existing grids has become an important operational challenge. In this paper, data center load accommodation (DCLA) refers to the maximum additional DC capacity that can be connected and reliably operated through coordinated scheduling of computing workloads and power-system resources without violating operating constraints.

Unlike conventional inflexible loads, DC demand is computing-driven and exhibits workload-dependent temporal flexibility. Inference, fine-tuning, and training workloads have different degrees of deferrability and interruptibility, allowing some workloads to be shifted or temporarily suspended to relieve network congestion~\cite{chen2021_idc_load_model,colangelo2025_grid_interactive}. However, such flexibility is bounded by quality-of-service (QoS) requirements, including service continuity and workload completion. Moreover, long-duration training workloads rely on checkpointing to save training states for recovery, and synchronized checkpoint operations across large clusters can cause rapid and substantial load releases that threaten frequency security (FS)~\cite{vercellino2026_ai_power,choukse2025_power_stabilization}. Workload flexibility therefore has a dual effect: temporal shifting can improve DCLA, whereas synchronized training operations can increase frequency-security risks. Consequently, grid-feasible DC accommodation may still be operationally infeasible because of QoS violations or unacceptable frequency excursions.

First, existing load-accommodation studies consider network constraints and operational uncertainty~\cite{madavan2024_risk_hca}, as well as dynamic operating limits~\cite{qin2026_operational_hc}. Flexible grid resources, such as energy storage (ES) and soft open points, have also been used to improve load accommodation capability~\cite{diao2025_msop_ess_hc}. These studies identify grid bottlenecks and quantify grid-side accommodation limits. However, they generally model additional loads as inflexible demand or exogenously specified flexibility, and therefore do not capture the service feasibility or checkpoint-induced frequency impacts of flexible DC workloads.

Second, DC flexibility studies have developed workload-to-power models and demand-response formulations~\cite{chen2021_idc_load_model,radovanovic2022_power_modeling}. Recent studies further couple computing workloads with power-system regulation and operation~\cite{li2024_computation_power,zhou2025_dc_load_modeling}, while market-oriented frameworks optimize flexible DC operation under electricity-market incentives~\cite{cui2026_fdc_market}. These studies demonstrate the operational value of workload flexibility, but most assume predetermined DC capacity rather than determining the maximum accommodable capacity. Moreover, representing flexibility using a fixed shiftable fraction or only a daily energy constraint may allow excessive workload reshaping or implicit curtailment, thereby obscuring the effect of QoS requirements on DCLA.

Third, frequency-security studies have highlighted the risks posed by fast variations in emerging large loads~\cite{nerc_large_loads2025,nerc_gaps_2026}. Recent measurements further show that AI training workloads can exhibit synchronized power variations and checkpoint-related transients~\cite{choukse2025_power_stabilization,vercellino2026_ai_power}. Frequency-secure scheduling models incorporate rate of change of frequency (RoCoF), frequency-peak, and quasi-steady-state (QSS) limits~\cite{badesa2019_frequency_services}. However, disturbance magnitude is typically treated as exogenous. Existing formulations therefore do not endogenously link checkpoint-induced power releases to accommodated training capacity, workload composition, scheduled throughput, and synchronization level, nor quantify their impact on maximum DCLA.

Taken together, existing studies do not capture the endogenous chain linking DC accommodation, workload scheduling, checkpoint-induced power releases, and frequency security. This is important because greater workload flexibility does not necessarily imply greater secure accommodation: the grid benefits of increased deferrability may be offset by checkpoint-induced frequency-security limits. Therefore, DCLA cannot be determined solely by grid adequacy.

To address these gaps, this paper develops a DCLA method that jointly accounts for QoS-constrained workload flexibility, checkpoint-induced frequency-security risks, and grid adequacy. The main contributions are summarized as follows:

\begin{figure}[!t]
\centering
\includegraphics[width=0.95\linewidth]{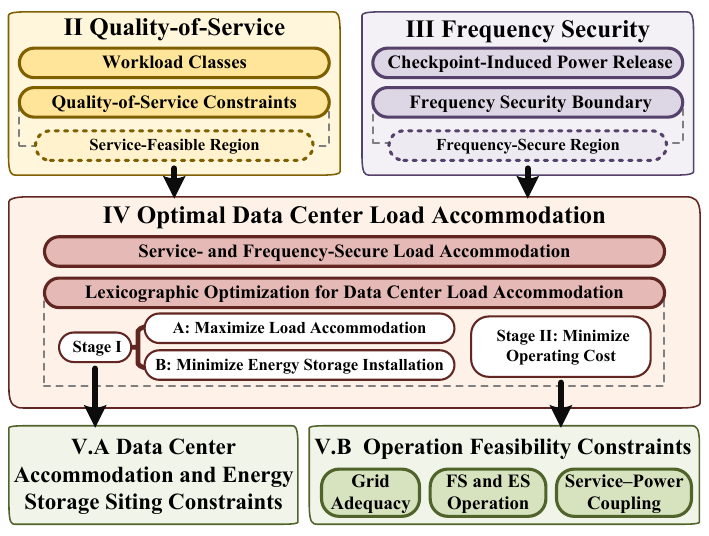}
\caption{Overall structure of the proposed DCLA methodology integrating workload QoS, checkpoint-induced frequency security, grid adequacy, and lexicographic optimization.}
\label{fig:overall}
\end{figure}

\begin{itemize}

\item QoS constraints are formulated for flexible DC workloads in DCLA. Workloads are classified according to latency sensitivity, deferrability, and interruptibility. Rolling-window throughput continuity and daily workload completion are used to ensure service feasibility and prevent implicit workload curtailment. These constraints quantify the reduction in DCLA caused by service requirements.

\item Checkpoint-induced FS constraints are formulated by endogenously linking periodic and event-triggered power releases to accommodated training workloads, scheduled throughput, checkpoint power-drop ratios, and synchronization levels. Using a reduced-order frequency-response model, the RoCoF, frequency-peak, and QSS limits are reformulated as algebraic constraints compatible with DCLA optimization.

\item A QoS- and FS-aware DCLA formulation is proposed to quantify the maximum DC load that can be accommodated under joint grid-adequacy, QoS, and FS constraints. A two-stage lexicographic formulation is developed. Stage I maximizes QoS- and FS-aware DCLA and minimizes the required ES deployment, while Stage II fixes the siting and sizing decisions and minimizes the weighted multi-stakeholder operating objective. Case studies show that QoS is the primary limiting factor, while frequency-security constraints become more restrictive as training workloads increase, demonstrating that greater workload flexibility does not necessarily increase secure DCLA.

\end{itemize}

The overall relationship among the main components of the proposed methodology is illustrated in Fig.~\ref{fig:overall}. Sections II and III establish the QoS constraints and checkpoint-induced FS constraints, respectively. These constraints are integrated with grid adequacy in Section IV to define QoS- and FS-aware DCLA, compare it with Grid-Centric DCLA, and formulate the lexicographic objectives. Section V presents the complete optimization constraints. Section VI provides case studies, and Section VII concludes the paper.

\section{Quality-of-Service Constraints of Flexible DC Workloads}

This section quantifies service feasibility for flexible DC workloads. Workloads are first classified according to their service flexibility, and the corresponding QoS constraints are then formulated for the DCLA problem.

\subsection{Workload Flexibility Classification}

From the grid perspective, a DC is a large nodal load subject to connection capacity and network security limits. However, its accommodable capacity depends not only on grid adequacy, but also on whether its computing demand can be deferred or interrupted. Workload flexibility must therefore be explicitly represented in the DCLA formulation.

\begin{figure}[!t]
\centering

\begin{minipage}[t]{0.44\columnwidth}
\vspace{0pt}
\centering
\includegraphics[width=1\linewidth]{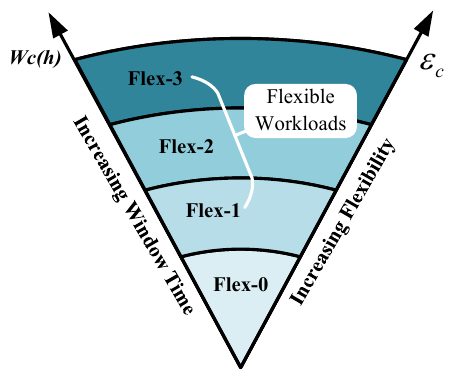}
\caption{Workload flexibility hierarchy from Flex-0 to Flex-3.}
\label{fig:fan_flexibility}
\end{minipage}
\hfill
\begin{minipage}[t]{0.54\columnwidth}
\vspace{0pt}
\centering
\captionof{table}{Rolling-Window Quality of Service Requirements}
\label{tab:task_parameters}
\scriptsize
\renewcommand{\arraystretch}{1.2}
\setlength{\tabcolsep}{3pt}
\begin{tabularx}{\linewidth}{@{}lX@{}}
\hline\hline
\textbf{Class} &
\textbf{Rolling-Window Requirement} \\
\hline
Flex-0 & No service deferral or shortfall. \\
Flex-1 & At most 10\% service shortfall over 3~h. \\
Flex-2 & At most 25\% service shortfall over 5~h. \\
Flex-3 & At most 50\% service shortfall over 6~h. \\
\hline\hline
\end{tabularx}
\par\vspace{2pt}
\parbox{\linewidth}{\scriptsize\textit{Note:} Flex-0 is rigid. Larger class numbers indicate longer allowable deferral and larger temporary service shortfalls.}
\end{minipage}

\end{figure}

We adopt a four-level workload classification based on latency sensitivity, deferrability, and interruptibility~\cite{colangelo2025_grid_interactive,epri_dcflex2025}. Flex-0 represents latency-critical and non-interruptible services, such as real-time inference and online search. These workloads require immediate execution and have negligible temporal elasticity, and are therefore modeled as rigid loads~\cite{liu2012_workload_management}. Flex-1 to Flex-3 exhibit progressively greater deferrability and interruptibility. Flex-1 represents batch inference that can be deferred over a short time window. Flex-2 represents fine-tuning workloads that can be temporarily suspended, whereas Flex-3 represents long-duration training workloads with the greatest temporal shifting capability~\cite{hu2021_gpu_workloads}. The resulting flexibility hierarchy is illustrated in Fig.~\ref{fig:fan_flexibility}.

To embed DC service requirements into system-level DCLA optimization, this qualitative hierarchy is translated into class-dependent rolling-window service envelopes that provide an aggregate representation of fine-grained SLA characteristics and bound short-term deviations from the reference execution profile.
As summarized in Table~\ref{tab:task_parameters}, the rolling-window lengths and allowable service shortfalls are parameterized based on the workload-flexibility characteristics and demonstrated service levels reported in~\cite{colangelo2025_grid_interactive,epri_dcflex2025}, with more flexible workloads allowing a larger temporary service shortfall over a longer time window. The shifted workload volume is preserved through the daily workload completion constraint.

\subsection{Quality-of-Service Constraints}

The service-side feasible region is described by rolling-window throughput continuity and daily volume conservation. Here, $n$, $c$, and $t$ index candidate DC buses, workload classes, and scheduling periods, respectively; $\Omega$ is the set of candidate DC buses; $\mathcal{C}=\{0,1,2,3\}$ is the set of workload classes; and $T$ is the number of scheduling periods. The parameter $d_{n,c,t}$ represents the reference execution profile of an admitted workload pool whose jobs and required inputs are available within the scheduling horizon; it does not represent a time-dependent job-arrival process. A relaxed daily adequacy constraint is also introduced for the Grid-Centric DCLA benchmark.

\subsubsection{Rolling-Window Throughput Continuity}

Rolling-window throughput continuity limits short-term deviations from the reference execution profile. The delivered throughput must remain above a prescribed fraction of the reference workload volume in every rolling window:
\begin{equation}
\label{eq:qos_window}
\begin{aligned}
&\sum_{\tau=t}^{t+W_c-1} \Gamma_{n,c,\tau} 
\geq 
(1-\epsilon_c) 
\sum_{\tau=t}^{t+W_c-1} d_{n,c,\tau},\\
&\quad \forall n\in\Omega,\ \forall c,\ 
\forall t=1,\dots,T-W_c+1 .
\end{aligned}
\end{equation}
Here, $\Gamma_{n,c,\tau}$ is the delivered throughput of workload class $c$ at bus $n$ in scheduling period $\tau$, where $\tau$ is the period index within a rolling window. The parameters $W_c$ and $\epsilon_c$ are the rolling-window length and allowable shortfall ratio of class $c$, respectively.
Constraint~\eqref{eq:qos_window} allows controlled advancement and deferral relative to the reference execution profile, but prevents sustained under-service within a short horizon.

\subsubsection{Daily Workload Completion}

Daily volume conservation ensures that flexibility is achieved by shifting workloads rather than curtailing them:
\begin{equation}
\label{eq:qos_daily}
\sum_{t=1}^{T} \Gamma_{n,c,t} 
= 
\sum_{t=1}^{T} d_{n,c,t},
\quad \forall n\in\Omega,\ \forall c .
\end{equation}
Constraint~\eqref{eq:qos_daily} guarantees that the total admitted workload represented by the reference execution profile is completed within the scheduling horizon. Together, \eqref{eq:qos_window} and \eqref{eq:qos_daily} define the service-side feasible region.

\subsubsection{Relaxed Daily Adequacy Benchmark}

For comparison, the Grid-Centric DCLA benchmark adopts the relaxed daily adequacy constraint
\begin{equation}
\label{eq:qos_daily_flex}
\sum_{t=1}^{T} \Gamma_{n,c,t}
\geq
(1-\epsilon_c)
\sum_{t=1}^{T} d_{n,c,t},
\quad \forall n\in\Omega,\ \forall c .
\end{equation}
Here, workload flexibility is approximated by the fixed parameter $\epsilon_c$, which specifies the maximum fraction of reference workload not required to be completed within the scheduling horizon. The remaining workload may be reshaped across periods subject to grid constraints. This approximation captures intraday flexibility but does not enforce
rolling-window continuity or full daily completion, and is therefore used only as a relaxed upper-bound benchmark.

\section{Checkpoint-Induced Frequency Security Boundary}

This section establishes a checkpoint-induced FS boundary for DCLA optimization. The checkpoint-induced power release is first quantified, and the RoCoF, frequency-peak, and QSS limits are then derived to constrain the resulting overfrequency risk.

\subsection{Checkpoint-Induced Power Release}

\begin{figure}[!t]
\centering
\includegraphics[width=0.8\columnwidth]{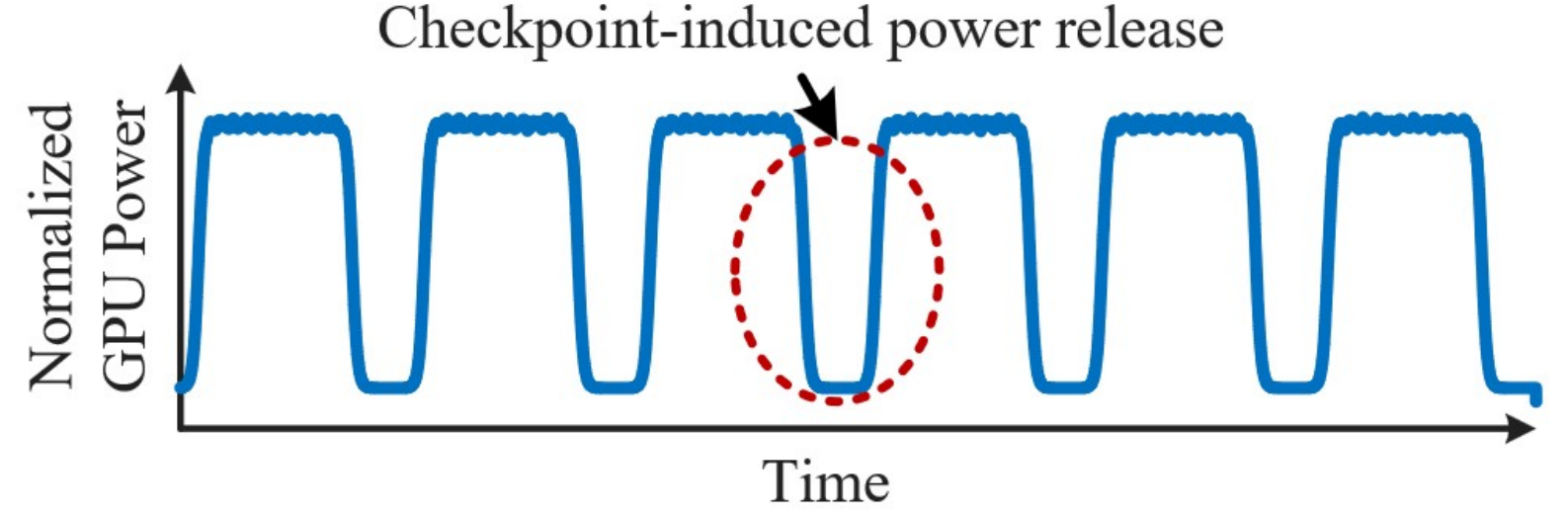}
\caption{Checkpoint-induced GPU power variation during AI training. Each synchronized checkpoint produces a rapid power drop, which appears to the grid as a short-duration load release and can create an overfrequency disturbance.}
\label{fig:checkpoint_power_profile}
\end{figure}

Checkpointing introduces synchronized transient phases in large-scale training and fine-tuning workloads. When many graphics processing units (GPUs) or training jobs enter the checkpoint phase simultaneously, the DC load may change rapidly, producing short-term power ramps and frequency security concerns~\cite{nerc_large_loads2025,choukse2025_power_stabilization}.
As illustrated in Fig.~\ref{fig:checkpoint_power_profile}, synchronized checkpoint operations can cause repeated and rapid reductions in GPU power.

This paper considers two checkpointing modes. Periodic checkpointing saves training states at predefined intervals to support fault tolerance and recovery after unexpected interruptions. Event-triggered checkpointing saves states in response to operational actions, such as task suspension, power curtailment, or GPU reallocation, so that the workload can later resume from the saved state. The recovery process after task reloading may cause a reverse load ramp, which can be mitigated by smooth recovery or snapback prevention~\cite{colangelo2025_grid_interactive}. 
This reverse ramp is not treated as the main constraint in this paper.

\subsubsection{Periodic Checkpoint-Induced Power Release}

The equivalent power release caused by periodic checkpointing is
\begin{equation}
\label{eq:cp_periodic}
\begin{aligned}
\Delta p_{n,t}^{\mathrm{per}}
&=
h^{\mathrm{dc}}
\sum_{c\in\mathcal{C}^{\mathrm{tr}}}
\alpha_{c,t}^{\mathrm{per}}
\Gamma_{n,c,t}, \\
\alpha_{c,t}^{\mathrm{per}}
&=
s_c^{\rho}s_c^{\eta}
\rho_t^{\mathrm{per}}\eta_t^{\mathrm{per}} .
\end{aligned}
\end{equation}
Here, $\Delta p_{n,t}^{\mathrm{per}}$ is the periodic checkpoint-induced power release at bus $n$ in period $t$, $h^{\mathrm{dc}}$ is the rated grid-facing power of one DC block, and $\mathcal{C}^{\mathrm{tr}}\subseteq\mathcal{C}$ is the set of checkpoint-enabled workload classes. The coefficient $\alpha_{c,t}^{\mathrm{per}}$ represents the periodic disturbance intensity of class $c$. The factors $s_c^{\rho}$ and $s_c^{\eta}$ are the class-dependent power-drop and synchronization factors, while $\rho_t^{\mathrm{per}}$ and $\eta_t^{\mathrm{per}}$ are the periodic power-drop and synchronization profiles.

\subsubsection{Event-Triggered Checkpoint-Induced Power Release}

Similarly, the equivalent power release caused by event-triggered checkpointing is
\begin{equation}
\label{eq:cp_event}
\begin{aligned}
\Delta p_{n,t}^{\mathrm{ev}}
=
h^{\mathrm{dc}}
\sum_{c\in\mathcal{C}^{\mathrm{tr}}}
\alpha_{c,t}^{\mathrm{ev}}
\Gamma_{n,c,t},\quad 
\alpha_{c,t}^{\mathrm{ev}}
=
s_c^{\rho}s_c^{\eta}
\rho_t^{\mathrm{ev}}\eta_t^{\mathrm{ev}} .
\end{aligned}
\end{equation}
Here, $\Delta p_{n,t}^{\mathrm{ev}}$ is the event-triggered checkpoint-induced power release, $\alpha_{c,t}^{\mathrm{ev}}$ is the corresponding disturbance coefficient, and $\rho_t^{\mathrm{ev}}$ and $\eta_t^{\mathrm{ev}}$ are the event-triggered power-drop and synchronization profiles.

\subsubsection{Total Checkpoint-Induced Disturbance}

The total equivalent checkpoint disturbance is
\begin{equation}
\label{eq:cp_total}
\Delta p_{n,t}^{\mathrm{cp}}
=
\Delta p_{n,t}^{\mathrm{per}}
+
\chi_t\Delta p_{n,t}^{\mathrm{ev}}.
\end{equation}
Here, $\Delta p_{n,t}^{\mathrm{cp}}$ is the total checkpoint-induced disturbance at bus $n$, and $\chi_t$ is the event-triggered checkpoint activation level in period $t$.
Equations~\eqref{eq:cp_periodic}--\eqref{eq:cp_total} link frequency risk with workload size, checkpoint triggering, and synchronization level. They do not assume a fixed power reduction ratio for the whole DC.

\subsection{Frequency Security Boundary}

This subsection converts checkpoint-induced disturbances into algebraic FS constraints. For each scheduling interval, the scheduled operating point defines the pre-disturbance condition, while a credible checkpoint event is evaluated on the seconds-scale frequency trajectory. Thus, the RoCoF, frequency-peak, and QSS limits are embedded into each scheduling interval using the maximum credible short-term power release rather than interval-averaged energy.

Since synchronous-unit commitment is not optimized, the equivalent system inertia coefficient is treated as constant and calculated as
\begin{equation}
\label{eq:equivalent_inertia}
M^{\mathrm{eq}}
=
\frac{2H^{\mathrm{syn}}}{f_0}
\sum_{g\in\mathcal{G}^{\mathrm{syn}}}
\bar p_g .
\end{equation}
Here, $M^{\mathrm{eq}}$ is the equivalent system inertia coefficient, $H^{\mathrm{syn}}$ is the inertia constant assigned to online synchronous generators, $f_0$ is the nominal frequency, $\mathcal{G}^{\mathrm{syn}}$ is the set of online synchronous generators, $g$ indexes a generator, and $\bar p_g$ is its maximum output.

For the overfrequency event caused by checkpoint-induced load reduction, $P_t^{\mathrm{cp}}$ denotes the aggregated checkpoint-induced disturbance in scheduling period $t$, and $P_t^0$ represents zero-delay disturbance mitigation available at $u=0^{+}$, where $u$ is the continuous time after the disturbance. The residual disturbance is therefore
$P_t^r=P_t^{\mathrm{cp}}-P_t^0\geq 0$. If such zero-delay mitigation is unavailable, $P_t^0$ is set to zero.

During the frequency-peak stage, the transient frequency dynamics are written as
\begin{equation}
\label{eq:swing_checkpoint}
M^{\mathrm{eq}}
\frac{d\Delta f_t(u)}{du}
=
P_t^r
-
p_t^f(u)
-
D_t^{\mathrm{dc}}\Delta f_t(u),
\end{equation}
Here, $\Delta f_t(u)$ is the dynamic frequency deviation, $p_t^f(u)$ is the delivered fast-response trajectory, and $D_t^{\mathrm{dc}}$ is the aggregate DC frequency-damping coefficient. A positive $\Delta f_t(u)$ denotes an overfrequency deviation.
The fast response is represented by
\begin{equation}
\label{eq:fast_response_trajectory}
p_t^f(u)
=
R_t^f\frac{u}{T^f},
\qquad
0\leq u\leq T^f .
\end{equation}
Here, $R_t^f$ is the aggregated fast-response capacity and $T^f$ is its delivery time.

\subsubsection{RoCoF Limit}

At the initial instant, $\Delta f_t(0)=0$, and the ramp-type response has
not been delivered. Since $P_t^0$ represents zero-delay disturbance
mitigation available at $u=0^{+}$, the initial RoCoF is determined by the
residual disturbance and the equivalent inertia:
\begin{equation}
\label{eq:initial_rocof}
\left.
\frac{d\Delta f_t}{du}
\right|_{0^{+}}
=
\frac{P_t^r}{M^{\mathrm{eq}}}.
\end{equation}
The RoCoF limit is therefore
\begin{equation}
\label{eq:sof_rocof}
P_t^r
\le
M^{\mathrm{eq}}\bar r,
\quad \forall t .
\end{equation}
Here, $\bar r$ is the maximum allowable rate of change of frequency.

\subsubsection{Frequency-Peak Limit}

Following the no-damping ramp-response approximation
in~\cite{badesa2019_frequency_services}, the frequency peak occurs at
\begin{equation}
\label{eq:peak_time}
u_t^\star
=
\frac{P_t^rT^f}{R_t^f}.
\end{equation}
Here, $u_t^\star$ is the time of the frequency peak in scheduling period $t$.
The peak occurs within the fast-response delivery interval if
\begin{equation}
\label{eq:peak_validity}
R_t^f
\geq
P_t^r,
\quad \forall t .
\end{equation}
The frequency-peak limit can therefore be represented as the following rotated second-order cone constraint:
\begin{equation}
\label{eq:sof_peak}
\left(P_t^r\right)^2
\leq
\frac{2M^{\mathrm{eq}}\overline{\Delta f}}{T^f}
R_t^f,
\quad \forall t .
\end{equation}
Here, $\overline{\Delta f}$ is the maximum allowable frequency-peak deviation.

\subsubsection{QSS Frequency Limit}

In the QSS stage, the transient fast response is replaced
by sustained response, while the damping- and droop-based responses are
available. Since the considered event is a unidirectional overfrequency
disturbance, the sustained response is limited to avoid overcompensation:
\begin{equation}
\label{eq:qss_sustained_bound}
0
\leq
P_t^p
\leq
P_t^r,
\quad
\forall t .
\end{equation}
Here, $P_t^p$ is the aggregated sustained response in period $t$.

The QSS frequency deviation is
\begin{equation}
\label{eq:qss_deviation}
\Delta f_t^{\mathrm{ss}}
=
\frac{
P_t^r-P_t^p
}{
D_t^{\mathrm{dc}}+K^g+K^{\mathrm{ibr}}
}.
\end{equation}
Here, $\Delta f_t^{\mathrm{ss}}$ is the quasi-steady-state frequency deviation, while $K^g$ and $K^{\mathrm{ibr}}$ are the aggregate governor and inverter-based-resource (IBR) quasi-steady-state response coefficients, respectively.

The QSS frequency limit is therefore
\begin{equation}
\label{eq:sof_qss}
P_t^r-P_t^p
\leq
\left(
D_t^{\mathrm{dc}}+K^g+K^{\mathrm{ibr}}
\right)
\overline{\Delta f}^{\mathrm{ss}},
\quad
\forall t .
\end{equation}
Here, $\overline{\Delta f}^{\mathrm{ss}}$ is the maximum allowable quasi-steady-state frequency deviation.

Constraints~\eqref{eq:sof_rocof}, \eqref{eq:peak_validity},
\eqref{eq:sof_peak}, \eqref{eq:qss_sustained_bound},
and~\eqref{eq:sof_qss} form the algebraic FS boundary. The peak
constraint adopts a conservative single-response no-damping
approximation, while sustained and droop-based responses are credited in
the QSS constraint. The transition between the fast and
sustained response stages is assumed not to create a larger secondary
frequency excursion.

\section{Data Center Load Accommodation and Bias Quantification}

This section quantifies DCLA and its systematic bias. Based on the QoS constraints in Section~II and the FS boundary in Section~III, Grid-Centric DCLA and QoS- and FS-aware DCLA are first compared. Then, a two-stage lexicographic formulation is used to maximize the accommodable DC load, minimize the required ES installation, and minimize operating cost in sequence.

\subsection{Grid-Centric DCLA and QoS- and FS-Aware DCLA}

Traditional DCLA studies often treat DCs as rigid loads or fully dispatchable demand. Both assumptions can be misleading. A grid-feasible DC schedule may still be infeasible if it violates service continuity, leaves workloads unfinished, or creates unacceptable frequency security risk. Therefore, DCLA should be defined over the joint grid, service, and FS feasible regions.

\subsubsection{Grid-Centric DCLA}

The Grid-Centric DCLA feasible region $\Omega_g$ includes the grid-side
constraints and the relaxed daily adequacy constraint
\eqref{eq:qos_daily_flex}. The maximum Grid-Centric DCLA
is defined as

\begin{definition}[Grid-Centric DCLA]
\label{def:traditional}
\begin{equation}
\label{eq:Definition1}
H_{g}
:=
\max
\left\{
\sum_{n\in\Omega} H_n
\;:\;
\Xi \in \Omega_g
\right\}.
\end{equation}
\end{definition}
Here, $H_g$ is the maximum Grid-Centric DCLA, $H_n$ is the accommodated DC capacity at bus $n$, and $\Xi$ is the collection of accommodation and operating variables.

This benchmark excludes rolling-window continuity, full daily workload
completion, and the FS boundary. It may therefore admit schedules that
are grid-feasible but operationally infeasible.

\subsubsection{QoS- and FS-Aware DCLA}

The QoS- and FS-aware maximum accommodable DC load is defined as

\begin{definition}[QoS- and FS-Aware DCLA]
\label{def:qos}
\begin{equation}
\label{eq:Definition2}
H_{s} := 
\max 
\left\{ 
\sum_{n\in\Omega} H_n 
\;:\; 
\Xi \in \Omega_g \cap \Omega_s \cap \Omega_f 
\right\}.
\end{equation}
\end{definition}
Here, $H_s$ is the maximum QoS- and FS-aware DCLA, while $\Omega_s$ and $\Omega_f$ are the strict-service and frequency-security feasible regions, respectively.

Definition~\ref{def:qos} ensures that the accommodated load is feasible under grid adequacy, QoS, and frequency security. Since it imposes stricter operational requirements than Definition~\ref{def:traditional}, the Grid-Centric DCLA definition may overestimate the QoS- and FS-aware DCLA.

\subsubsection{Systematic Bias of Grid-Centric DCLA}

\begin{figure}[!t]
\centering
\includegraphics[width=0.7\linewidth]{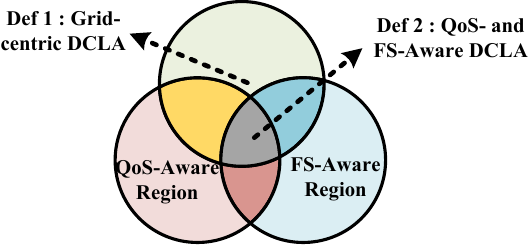}
\caption{Feasible-region interpretation of the two DCLA definitions. Grid-Centric DCLA is optimized over $\Omega_g$, whereas QoS- and FS-aware DCLA is restricted to $\Omega_g\cap\Omega_s\cap\Omega_f$; the difference between their maxima produces the systematic bias.}
\label{fig:feasible_regions_bias}
\end{figure}

As shown in Fig.~\ref{fig:feasible_regions_bias}, the systematic bias rate is defined as
\begin{equation}
\label{eq:bias}
\delta_b = \frac{H_{g} - H_{s}}{H_{g}}.
\end{equation}
A positive $\delta_b$ means that the Grid-Centric DCLA model overestimates the feasible DC load. A larger $\delta_b$ indicates that QoS or FS constraints become dominant factors limiting QoS- and FS-aware DCLA.

\subsection{Lexicographic Formulation of DCLA}
The objectives are ordered lexicographically. Stage I-A first maximizes the accommodated DC load, Stage I-B then minimizes the required number of installable ES units without reducing the maximum DCLA, and Stage II finally minimizes a multi-stakeholder operating objective without changing the preceding siting and sizing decisions.

\subsubsection{Stage I-A: Maximum DCLA}
The first optimization level maximizes the total accommodable DC load:
\begin{equation}
\label{eq:stage1a_obj}
\max\quad
\sum_{n\in\Omega}H_n .
\end{equation}
The resulting maximum DCLA is denoted by $H^\star$.

\subsubsection{Stage I-B: Minimum ES Installation}
After fixing the maximum DCLA, the second optimization level minimizes the required number of installable ES units:
\begin{equation}
\label{eq:stage1b_obj}
\begin{aligned}
\min\quad
&\sum_{n\in\Omega}x_n^{\mathrm{es}}\\
\mathrm{s.t.}\quad
&\sum_{n\in\Omega}H_n=H^\star .
\end{aligned}
\end{equation}
Here, $x_n$ is the number of accommodated DC blocks at bus $n$, and $x_n^{\mathrm{es}}$ is the number of installable energy storage units at bus $n$.
The resulting minimum ES requirement is denoted by $X_{\mathrm{es}}^\star$. The corresponding nodal DC and ES siting and sizing decisions are denoted by $x_n^\star$ and $x_n^{\mathrm{es},\star}$, respectively.

\subsubsection{Stage II: Multi-Stakeholder Operating Objective}

After fixing the DC and ES siting and sizing decisions obtained in Stage I-B, Stage II coordinates the interests of the grid operator and the DC operator through a weighted multi-stakeholder objective:
\begin{equation}
\label{eq:stage2_obj}
\begin{aligned}
\min\quad
&w^{\mathrm{grid}}
\sum_{t=1}^{T}
\sum_{g\in\mathcal G^{\mathrm d}}
C_g(p_{g,t})\Delta t\\
&+
w^{\mathrm{dc}}
\sum_{t=1}^{T}
\lambda_t
\sum_{n\in\Omega}
p_{n,t}^{\mathrm{dc}}
\Delta t\\
&+
c^{\mathrm{es}}
\sum_{t=1}^{T}
\sum_{n\in\mathcal E}
\left(
p_{n,t}^{\mathrm{ch}}
+
p_{n,t}^{\mathrm{dis}}
\right)
\Delta t ,
\end{aligned}
\end{equation}
Here, $w^{\mathrm{grid}}$ and $w^{\mathrm{dc}}$ are the weights of the grid-operator and DC-operator objectives; $\mathcal{G}^{\mathrm d}$ is the set of dispatchable generators; $C_g(\cdot)$ is the generation cost function; $p_{g,t}$ is the output of generator $g$; and $\Delta t$ is the duration of one scheduling period. The parameter $\lambda_t$ is the electricity price, $p_{n,t}^{\mathrm{dc}}$ is the grid-facing DC power, $c^{\mathrm{es}}$ is the energy storage cycling penalty coefficient, and $\mathcal E$ is the set of all energy storage buses. The variables $p_{n,t}^{\mathrm{ch}}$ and $p_{n,t}^{\mathrm{dis}}$ are the charging and discharging powers.
The first two terms represent generation cost and DC electricity expenditure. They are not treated as additive social costs because electricity payments are market transfers. The last term is a small ES cycling penalty that suppresses unnecessary charging and discharging.

The Stage-I siting and sizing decisions are fixed by
\begin{equation}
\label{eq:stage2_fix}
x_n=x_n^\star,
\qquad
x_n^{\mathrm{es}}=x_n^{\mathrm{es},\star},
\quad
\forall n\in\Omega .
\end{equation}
Stage II therefore optimizes generator dispatch, workload scheduling, renewable utilization, and ES operation without changing the maximum DCLA, minimum ES requirement, or their nodal allocations.

\section{Optimization Constraints of DCLA Under QoS, FS, and Grid Adequacy}

This section presents the complete DCLA optimization model under grid adequacy, QoS, FS, and ES constraints. It first defines the DC accommodation and ES siting decisions. It then formulates the operational constraints for grid dispatch, workload scheduling, frequency response, and ES operation.

\subsection{Accommodation and Siting Decisions}

This subsection defines the candidate-bus constraints for DCLA and installable ES siting. The installable ES may be sited at any bus in the candidate set $\Omega$ and is not required to be co-located with an accommodated DC. It includes the discrete DC block representation, local capacity limits, local ES installation limits, and the system-wide ES budget.

\subsubsection{DC Block Accommodation Capacity}

The accommodated DC capacity at each candidate bus is represented as an integer multiple of a standardized DC block:
\begin{equation}
\label{eq:dc_block}
H_n
=
h^{\mathrm{dc}}x_n,
\quad
x_n\in
\{0,1,\ldots,\bar x_n^{\mathrm{dc}}\},
\quad
\forall n\in\Omega .
\end{equation}
Here, $\bar x_n^{\mathrm{dc}}$ is the maximum number of accommodated DC blocks allowed at bus $n$.

The accommodated DC capacity cannot exceed the available interconnection capacity at each candidate bus:
\begin{equation}
\label{eq:dc_capacity_limit}
0
\leq
H_n
\leq
\bar H_n^{\mathrm{dc}},
\quad
\forall n\in\Omega .
\end{equation}
Here, $\bar H_n^{\mathrm{dc}}$ is the maximum DC capacity at bus $n$.

\subsubsection{Installable ES Siting}

The number of ES units installed at each candidate bus is restricted by the local installation limit:
\begin{equation}
\label{eq:es_local_limit}
x_n^{\mathrm{es}}
\in
\{0,1,\ldots,\bar x_n^{\mathrm{es}}\},
\quad
\forall n\in\Omega .
\end{equation}
Here, $\bar x_n^{\mathrm{es}}$ is the local upper limit on installable energy storage units at bus $n$.

The total number of ES units installed across all candidate buses is limited by the system-wide ES budget:
\begin{equation}
\label{eq:es_budget}
\sum_{n\in\Omega}
x_n^{\mathrm{es}}
\leq
x_{\mathrm{budget}}^{\mathrm{es}} .
\end{equation}
Here, $x_{\mathrm{budget}}^{\mathrm{es}}$ is the system-wide energy storage unit budget.

\subsection{Operational Feasibility Constraints}

Given the accommodation and ES siting decisions, these constraints ensure feasible grid dispatch, DC service delivery, frequency response, and ES operation in both optimization stages.

\subsubsection{Grid Adequacy Constraints}

Grid adequacy is enforced through nodal power balance, DC power flow, and transmission security constraints. DC and ES variables are set to zero at buses without the corresponding facilities.

At each bus and period, the total power injection must equal the net outgoing line flow. This constraint couples generator and renewable outputs with DC demand, ES operation, and baseline load:
\begin{equation}
\label{eq:balance}
\begin{aligned}
&\sum_{g\in G_n}p_{g,t}
+\sum_{w\in W_n}p_{w,t}^{\mathrm w}
+\sum_{s\in S_n}p_{s,t}^{\mathrm s}
+p_{n,t}^{\mathrm{dis}}
-p_{n,t}^{\mathrm{ch}}\\
&-p_{n,t}^{\mathrm{dc}}
-p_{n,t}^{L}
=
\sum_{k\in\delta(n)}f_{nk,t},
\quad
\forall n\in\mathcal{N},\ \forall t .
\end{aligned}
\end{equation}
Here, $\mathcal N$ is the set of buses, $k$ indexes a bus adjacent to bus $n$, and $\delta(n)$ is the set of buses adjacent to bus $n$. The sets $G_n$, $W_n$, and $S_n$ contain the dispatchable generators, wind units, and solar units at bus $n$, respectively, while $w$ and $s$ index wind and solar units. The variables $p_{w,t}^{\mathrm w}$ and $p_{s,t}^{\mathrm s}$ are the utilized wind and solar powers, $p_{n,t}^{L}$ is the non-DC baseline load, and $f_{nk,t}$ is the active power flow from bus $n$ to bus $k$.

The output of each dispatchable generator is bounded by its operating limits:
\begin{equation}
\label{eq:generator_output_limit}
\underline p_g
\leq
p_{g,t}
\leq
\bar p_g,
\quad
\forall g\in\mathcal{G}^{\mathrm d},\ \forall t .
\end{equation}
Here, $\underline p_g$ and $\bar p_g$ are the minimum and maximum outputs of generator $g$.

Its intertemporal output variation is limited by the upward and downward ramp rates:
\begin{equation}
\label{eq:generator_ramp_limit}
\begin{aligned}
&-R_g^{\mathrm{down}}\Delta t 
\leq p_{g,t}-p_{g,t-1} 
\leq R_g^{\mathrm{up}}\Delta t,\\ 
&\forall g\in\mathcal{G}^{\mathrm d}, \forall t=2,\ldots,T .
\end{aligned}
\end{equation}
Here, $R_g^{\mathrm{up}}$ and $R_g^{\mathrm{down}}$ are the upward and downward ramp-rate limits of generator $g$.

The utilized renewable power is limited by its available output:
\begin{equation}
\label{eq:renewable_output_limit}
\begin{aligned}
0 &\leq p_{w,t}^{\mathrm w}\leq \bar p_{w,t}^{\mathrm w},
&&\forall w,\ \forall t,\\
0 &\leq p_{s,t}^{\mathrm s}\leq \bar p_{s,t}^{\mathrm s},
&&\forall s,\ \forall t .
\end{aligned}
\end{equation}
Here, $\bar p_{w,t}^{\mathrm w}$ and $\bar p_{s,t}^{\mathrm s}$ are the available wind and solar powers.

The active power flow on each transmission line is represented by the DC power flow relation:
\begin{equation}
\label{eq:dc_pf}
f_{nk,t}
=
B_{nk}
\left(
\theta_{n,t}-\theta_{k,t}
\right),
\quad
\forall (n,k)\in\mathcal{L},\ \forall t .
\end{equation}
Here, $\mathcal L$ is the set of transmission lines, $B_{nk}$ is the susceptance of line $(n,k)$, and $\theta_{n,t}$ is the voltage angle at bus $n$.

The voltage angle at the reference bus is fixed:
\begin{equation}
\label{eq:reference_bus}
\theta_{n^{\mathrm{ref}},t}=0,
\quad
\forall t .
\end{equation}
Here, $n^{\mathrm{ref}}$ is the reference-bus index.

Under normal operating conditions, the flow on each monitored line must remain within its continuous thermal rating:
\begin{equation}
\label{eq:normal_line_limit}
-\bar f_{nk}^{A}
\leq
f_{nk,t}
\leq
\bar f_{nk}^{A},
\quad
\forall (n,k)\in\mathcal{L},\ \forall t .
\end{equation}
Here, $\bar f_{nk}^{A}$ is the continuous flow rating of line $(n,k)$.

Transmission security after a single-line outage is represented using line outage distribution factors~\cite{guo2009_lodf}. For each monitored line and contingency, the estimated post-contingency flow must remain within the emergency rating:
\begin{equation}
\label{eq:post_contingency_limit}
\begin{aligned}
&-\bar f_{nk}^{C}\leq
f_{nk,t}
+
\mathrm{L}_{nk}^{ij}f_{ij,t}
\leq
\bar f_{nk}^{C},\\
&\forall (n,k)\in\mathcal{L},
\forall (i,j)\in\mathcal{L}^{c},
\forall (i,j)\neq(n,k),\ \forall t .
\end{aligned}
\end{equation}
Here, $i$ and $j$ index the terminal buses of a contingency line, $\mathcal L^{c}$ is the set of considered line contingencies, $\mathrm{L}_{nk}^{ij}$ is the line outage distribution factor relating the outage of line $(i,j)$ to monitored line $(n,k)$, and $\bar f_{nk}^{C}$ is the emergency flow rating.

\subsubsection{DC Service and Power Coupling}

This part links the accommodated DC capacity, delivered workload throughput, and grid-facing DC power.

The delivered throughput of each workload class is limited by the computing capacity associated with the accommodated DC blocks:
\begin{equation}
\label{eq:dc_throughput_limit}
0
\leq
\Gamma_{n,c,t}
\leq
\bar\Gamma_{n,c}x_n,
\quad
\forall n\in\Omega,\
\forall c\in\mathcal{C},\
\forall t .
\end{equation}
Here, $\bar\Gamma_{n,c}$ is the per-block throughput limit of workload class $c$ at bus $n$.

The total throughput across all workload classes cannot exceed the aggregate computing capacity of the accommodated DC blocks:
\begin{equation}
\label{eq:dc_total_throughput_limit}
\sum_{c\in\mathcal C}\Gamma_{n,c,t}
\leq
\bar\Gamma^{\mathrm{dc}}x_n,
\quad
\forall n\in\Omega,\ \forall t .
\end{equation}
Here, $\bar\Gamma^{\mathrm{dc}}$ is the aggregate throughput limit of one DC block.

The grid-facing DC power includes idle information-technology (IT) power, workload-dependent IT power, and auxiliary power represented by power usage effectiveness (PUE)~\cite{chen2021_idc_load_model}:
\begin{equation}
\label{eq:dc_power_mapping}
p_{n,t}^{\mathrm{dc}}
=
\mathrm{PUE}
\left(
P_n^{\mathrm{idle}}x_n
+
\sum_{c\in\mathcal{C}}
e_{n,c}\Gamma_{n,c,t}
\right),
\quad
\forall n\in\Omega,\ \forall t .
\end{equation}
Here, $\mathrm{PUE}$ is the power usage effectiveness, $P_n^{\mathrm{idle}}$ is the idle IT power per DC block at bus $n$, and $e_{n,c}$ is the power-to-throughput coefficient of workload class $c$.

The grid-facing DC power cannot exceed the accommodated rated capacity at each candidate bus:
\begin{equation}
\label{eq:dc_power_limit}
0
\leq
p_{n,t}^{\mathrm{dc}}
\leq
H_n,
\quad
\forall n\in\Omega,\ \forall t .
\end{equation}

The reference execution profile of the admitted workload pool scales with the number of accommodated DC blocks:
\begin{equation}
\label{eq:base_demand_scaling}
d_{n,c,t}
=
x_n\bar\Gamma^{\mathrm{dc}}\tilde d_{c,t},
\quad
\forall n\in\Omega,\
\forall c\in\mathcal{C},\
\forall t .
\end{equation}
Here, $\tilde d_{c,t}$ is the normalized reference execution profile of workload class $c$.

For QoS- and FS-aware DCLA, these coupling constraints are combined with the rolling-window throughput continuity constraint in \eqref{eq:qos_window} and the daily workload completion constraint in \eqref{eq:qos_daily}. The Grid-Centric DCLA benchmark instead uses the relaxed daily adequacy constraint in \eqref{eq:qos_daily_flex}. This distinction separates feasible workload shifting from implicit service curtailment.

\subsubsection{Checkpoint and Frequency Security Coupling}

This part links the node-level workload decisions to the system-level checkpoint disturbance and frequency response quantities.

The total checkpoint-induced disturbance is obtained by aggregating the checkpoint power release across all candidate DC buses:
\begin{equation}
\label{eq:cp_aggregation}
P_t^{\mathrm{cp}}
=
\sum_{n\in\Omega}
\Delta p_{n,t}^{\mathrm{cp}},
\quad
\forall t .
\end{equation}

The immediate, fast, and sustained DC-side responses are aggregated across candidate buses:
\begin{equation}
\label{eq:response_aggregation}
\begin{aligned}
P_t^0
&=
\sum_{n\in\Omega}p_{n,t}^{0},\quad 
R_t^f
=
\sum_{n\in\Omega}r_{n,t}^{f},\\
P_t^p
&=
\sum_{n\in\Omega}p_{n,t}^{p},
\quad
\forall t .
\end{aligned}
\end{equation}
Here, $p_{n,t}^{0}$, $r_{n,t}^{f}$, and $p_{n,t}^{p}$ are the zero-delay mitigation, fast response, and sustained response allocated at candidate bus $n$, respectively.

After the zero-delay disturbance mitigation is applied, the remaining
disturbance is represented by
\begin{equation}
\label{eq:residual_disturbance}
P_t^r=P_t^{\mathrm{cp}}
-P_t^0,
\qquad
P_t^r
\geq
0,
\quad
\forall t .
\end{equation}

The available DC-side response capacity depends on the scheduled throughput of flexible workloads:
\begin{equation}
\label{eq:dc_response_capacity}
\bar P_{n,t}^{\mathrm{resp}}
=
h^{\mathrm{dc}}
\sum_{c\in\mathcal{C}}
\omega_c\Gamma_{n,c,t},
\quad
\forall n\in\Omega,\ \forall t .
\end{equation}
Here, $\bar P_{n,t}^{\mathrm{resp}}$ is the available DC-side response capacity, and $\omega_c$ is the response availability weight of workload class $c$.

The aggregate DC-side damping coefficient is determined by the accommodated Flex-1--Flex-3 workloads:
\begin{equation}
\label{eq:dc_damping_model}
D_t^{\mathrm{dc}}
=
\kappa_D
\frac{h^{\mathrm{dc}}}{f_0}
\sum_{n\in\Omega}
x_n\bar\Gamma^{\mathrm{dc}}
\sum_{c=1}^{3}
\tilde d_{c,t},
\quad
\forall t .
\end{equation}
Here, $\kappa_D$ is the DC frequency-damping scaling factor.

The zero-delay mitigation at each candidate bus is limited by its
available capacity and cannot exceed the local checkpoint disturbance:
\begin{equation}
\label{eq:immediate_response_limit}
\begin{aligned}
&0
\leq
p_{n,t}^{0}
\leq
\bar p_{n,t}^{0},\\
&p_{n,t}^{0}
\leq
\Delta p_{n,t}^{\mathrm{cp}},
\quad
\forall n\in\Omega,\ \forall t .
\end{aligned}
\end{equation}
Here, $\bar p_{n,t}^{0}$ is the available zero-delay mitigation capacity at bus $n$.

The allocated fast response is limited by a fixed fraction of the
available DC response capacity:
\begin{equation}
\label{eq:fast_response_limit}
0
\leq
r_{n,t}^{f}
\leq
\gamma^f
\bar P_{n,t}^{\mathrm{resp}},
\quad
\forall n\in\Omega,\ \forall t .
\end{equation}
Here, $\gamma^f$ is the fraction of available DC response capacity assigned to fast response.

The allocated sustained response is limited by a fixed fraction of the available DC response capacity:
\begin{equation}
\label{eq:sustained_response_limit}
0
\leq
p_{n,t}^{p}
\leq
\gamma^p
\bar P_{n,t}^{\mathrm{resp}},
\quad
\forall n\in\Omega,\ \forall t .
\end{equation}
Here, $\gamma^p$ is the fraction of available DC response capacity assigned to sustained response.

The aggregated disturbance and response quantities are constrained by
\eqref{eq:sof_rocof}, \eqref{eq:peak_validity},
\eqref{eq:sof_peak}, \eqref{eq:qss_sustained_bound},
and~\eqref{eq:sof_qss}.

The QSS contribution from IBRs cannot exceed their available response headroom:
\begin{equation}
\label{eq:ibr_support_model}
K^{\mathrm{ibr}}
\overline{\Delta f}^{\mathrm{ss}}
\leq
\bar P_t^{\mathrm{ibr}},
\quad
\forall t .
\end{equation}
Here, $\bar P_t^{\mathrm{ibr}}$ is the available inverter-based-resource response headroom.

These constraints make the FS boundary dependent on workload composition, scheduled throughput, and checkpoint synchronization rather than on a fixed DC power-drop ratio.

\subsubsection{ES Operation Constraints}

ES provides intertemporal energy buffering to relieve network congestion
and renewable power imbalances~\cite{gayme2013_storage}. All ES units use the
same multi-period operating model.

At candidate DC buses, the available ES capacities are coupled with the
installation decisions:
\begin{equation}
\label{eq:es_installed_capacity}
\bar p_n^{\mathrm E}
=
p^{\mathrm{es}}x_n^{\mathrm{es}},
\qquad
\bar e_n^{\mathrm E}
=
e^{\mathrm{es}}x_n^{\mathrm{es}},
\quad
\forall n\in\Omega .
\end{equation}
Here, $\mathcal R$ is the set of buses with fixed energy storage; $\bar p_n^{\mathrm E}$ and $\bar e_n^{\mathrm E}$ are the available energy storage power and energy capacities; and $p^{\mathrm{es}}$ and $e^{\mathrm{es}}$ are the power and energy ratings of one installable energy storage unit. At buses in $\mathcal R$, $\bar p_n^{\mathrm E}$ and $\bar e_n^{\mathrm E}$ are fixed.

The ES energy state evolves as
\begin{equation}
\label{eq:es_energy_balance}
e_{n,t}
=
e_{n,t-1}
+
\eta^{\mathrm{ch}}p_{n,t}^{\mathrm{ch}}\Delta t
-
\frac{p_{n,t}^{\mathrm{dis}}}{\eta^{\mathrm{dis}}}\Delta t,
\quad
\forall n\in\mathcal E,\ \forall t .
\end{equation}
Here, $e_{n,t}$ is the stored energy, while $\eta^{\mathrm{ch}}$ and $\eta^{\mathrm{dis}}$ are the charging and discharging efficiencies.

The ES energy state and charging and discharging powers satisfy
\begin{equation}
\label{eq:es_capacity_limits}
\begin{aligned}
0
&\leq
e_{n,t}
\leq
\bar e_n^{\mathrm E},\\
0
&\leq
p_{n,t}^{\mathrm{ch}}
\leq
\bar p_n^{\mathrm E},\\
0
&\leq
p_{n,t}^{\mathrm{dis}}
\leq
\bar p_n^{\mathrm E},
\end{aligned}
\quad
\forall n\in\mathcal E,\ \forall t .
\end{equation}

Simultaneous charging and discharging are prevented by the binary
operating mode variable:
\begin{equation}
\label{eq:es_operating_mode}
\begin{aligned}
p_{n,t}^{\mathrm{ch}}
&\leq
p_n^{\mathrm{E,max}}z_{n,t}^{\mathrm{es}},\\
p_{n,t}^{\mathrm{dis}}
&\leq
p_n^{\mathrm{E,max}}
\left(
1-z_{n,t}^{\mathrm{es}}
\right),\\
z_{n,t}^{\mathrm{es}}
&\in
\{0,1\},
\end{aligned}
\quad
\forall n\in\mathcal E,\ \forall t .
\end{equation}
Here, $z_{n,t}^{\mathrm{es}}$ is the binary charging-mode variable, and $p_n^{\mathrm{E,max}}$ is the big-M power bound used in the operating-mode constraint. For candidate DC buses, $p_n^{\mathrm{E,max}}=p^{\mathrm{es}}\bar x_n^{\mathrm{es}}$; for buses with fixed ES, it equals the fixed ES power capacity.

The initial and terminal energy states satisfy
\begin{equation}
\label{eq:es_boundary_states}
e_{n,0}
=
\beta^0\bar e_n^{\mathrm E},
\qquad
e_{n,T}
=
e_{n,0},
\quad
\forall n\in\mathcal E .
\end{equation}
Here, $\beta^0$ is the initial and terminal energy ratio.

\section{Case Studies}

This section evaluates the proposed DCLA method on the IEEE 39-bus system. The studies quantify the Grid-Centric DCLA bias and examine the effects of workload composition, workload flexibility, installable ES, and FS constraints.

\subsection{Case Setup}

The study uses the IEEE 39-bus system over a 24-h horizon (96 15-min intervals) with system demand varying between 75\% and 100\% of the nominal load. Wind and solar generation replace synchronous units at buses 38 and 37, respectively, following Electric Reliability Council of Texas (ERCOT) profiles~\cite{ercot_wind_actual,ercot_solar_actual}. Dispatchable generators have hourly ramping limits equal to 20\% of their operating range. Grid-side fixed 4-h ES is installed at buses 37 and 38, rated at 25\% of the connected renewable capacity. For installable ES, each candidate bus allows up to five 5 MW/20 MWh units~\cite{cole2023_battery}, capped at a system-wide budget of 20 units. These units are grid-connected resources modeled over the candidate-bus set and are not required to be co-located with accommodated DC capacity. All ES units operate with $\eta^{\mathrm{ch}}=\eta^{\mathrm{dis}}=\sqrt{0.85}$, a 50\% initial and terminal state of charge, and strictly non-simultaneous charging and discharging.
Although operating decisions are scheduled at a 15-min resolution, the FS constraints evaluate a seconds-scale checkpoint event at every 15-min operating point through the continuous-time frequency-response model.

Candidate DC buses are $\Omega=\{16,18,19,24,27\}$, each permitting up to fifteen 10 MW blocks with $\mathrm{PUE}=1.21$~\cite{eavms2025}. The IT-power function is parameterized using NVIDIA A100 GPU measurements~\cite{nvidia_a100,colangelo2025_grid_interactive}. Four workload classes are evaluated using inference-derived profiles for Flex-0 and Flex-1 and training-derived profiles for Flex-2 and Flex-3~\cite{burstgpt2024,choukse2025_power_stabilization}. Table~\ref{tab:workload_mix} lists their QoS parameters and shares across the balanced baseline, inference-heavy, and training-heavy mixes. Flex-2 and Flex-3 produce checkpoint-induced power disturbances with coefficients $s_c^\rho s_c^\eta$ of 0.48 and 1.00, respectively, whereas Flex-0 and Flex-1 do not. The maximum event-triggered activation factor is 0.8, and fast and sustained workload responses equal 60\% and 30\% of the weighted flexible-load response, respectively.

\begin{table}[!t]
\caption{Workload Mixes and QoS Settings}
\label{tab:workload_mix}
\centering
\footnotesize
\renewcommand{\arraystretch}{1.2}
\setlength{\tabcolsep}{4pt}
\begin{tabular}{@{}lcccc@{}}
\hline\hline
\textbf{Setting}
& \textbf{Flex-0}
& \textbf{Flex-1}
& \textbf{Flex-2}
& \textbf{Flex-3} \\
\hline
Balanced & 0.30 & 0.15 & 0.15 & 0.40 \\
Inference-heavy & 0.65 & 0.25 & 0.05 & 0.05 \\
Training-heavy & 0.10 & 0.05 & 0.15 & 0.70 \\
$W_c$ (periods) & 1 & 12 & 20 & 24 \\
$\epsilon_c$ & 0 & 0.10 & 0.25 & 0.50 \\
$\omega_c$ & 0 & 0.10 & 0.25 & 0.50 \\
\hline\hline
\end{tabular}
\par\vspace{2pt}
\parbox{\linewidth}{\scriptsize\textit{Note:} Workload shares in each mix sum to one. $W_c$ is measured in 15-min periods; $\epsilon_c$ is the allowable rolling-window service shortfall; and $\omega_c$ is the response-availability weight.}
\end{table}

Frequency-support parameters $M^{\mathrm{eq}}$ and $K^g$ are based on online synchronous units, assuming a 5\% governor droop and a 3\% available primary-response fraction. The DC damping coefficient is $\kappa_D=0.5$. Zero-delay mitigation and IBR-based QSS support are omitted. Table~\ref{tab:sof_parameters} details the main FS parameters~\cite{badesa2019_frequency_services}.

\begin{figure}[!t]
\centering

\begin{minipage}[t]{0.49\columnwidth}
\vspace{0pt}
\centering
\captionof{table}{FS Parameter Settings}
\label{tab:sof_parameters}
\scriptsize
\renewcommand{\arraystretch}{1.2}
\setlength{\tabcolsep}{1.5pt}
\begin{tabularx}{\linewidth}{
@{}
>{\centering\arraybackslash}X
>{\centering\arraybackslash}X
>{\centering\arraybackslash}X
>{\centering\arraybackslash}X
@{}}
\hline\hline
\textbf{Para.}
& \textbf{Value}
& \textbf{Para.}
& \textbf{Value} \\
\hline
$f_0$& 60 Hz& $H^{\mathrm{syn}}$& 4 s \\
$\bar r$& 0.5 Hz/s& $\overline{\Delta f}$& 0.8 Hz \\
$\overline{\Delta f}^{\mathrm{ss}}$& 0.5 Hz& $T^f$& 10 s \\
$\kappa_D$& 0.5& $K^{\mathrm{ibr}}$& 0 \\
$P_t^0$& 0& $\gamma^f$& 0.60 \\
$\gamma^p$& 0.30& $\max_t\chi_t$& 0.80 \\
\hline\hline
\end{tabularx}
\par\vspace{2pt}
\parbox{\linewidth}{\scriptsize\textit{Note:} The limits are evaluated at every 15-min operating point using the seconds-scale frequency-response model. $P_t^0=0$ and $K^{\mathrm{ibr}}=0$ mean that zero-delay mitigation and IBR-based QSS support are not credited.}
\end{minipage}
\hfill
\begin{minipage}[t]{0.49\columnwidth}
\vspace{0pt}
\centering
\captionof{table}{Baseline DCLA and ES Siting}
\label{tab:siting_results}
\scriptsize
\renewcommand{\arraystretch}{1.2}
\setlength{\tabcolsep}{2pt}
\begin{tabularx}{\linewidth}{
@{}
>{\centering\arraybackslash}X
>{\centering\arraybackslash}X
>{\centering\arraybackslash}X
@{}}
\hline\hline
\textbf{Bus}
& \textbf{Load (MW)}
& \textbf{ES Units} \\
\hline
16 & 130 & 0 \\
18 & 0 & 5 \\
19 & 150 & 5 \\
24 & 150 & 3 \\
27 & 0 & 5 \\
Total & 430 & 18 \\
\hline\hline
\end{tabularx}
\par\vspace{2pt}
\parbox{\linewidth}{\scriptsize\textit{Note:} Load denotes accommodated DC capacity. Each DC block is 10~MW and each installable ES unit is rated 5~MW/20~MWh. The allocation is one minimum-unit solution obtained after fixing the 430~MW maximum DCLA.}
\end{minipage}

\end{figure}

The lexicographic procedure follows Section~IV. Stage II uses ERCOT real-time prices in the DC-side electricity-expenditure term and applies equal stakeholder weights, $w^{\mathrm{grid}}=w^{\mathrm{dc}}=1$~\cite{ercot_prices}. The model is implemented in MATLAB using YALMIP and solved with Gurobi.

\subsection{Baseline QoS- and FS-Aware DCLA Results}

The complete QoS- and FS-aware formulation yields 430~MW of DCLA, corresponding to 43 DC blocks. Stage I-B requires 18 installable ES units. These results define the baseline for the following comparisons.

\subsubsection{Spatial Siting and ES Allocation}

Table~\ref{tab:siting_results} shows that buses 19 and 24 reach their 150 MW limits, bus 16 accommodates 130~MW, and buses 18 and 27 host no DC capacity.
The ES allocation follows a different pattern. Buses 18, 19, 24, and 27 host ES, while bus 16 supports 130~MW of DC capacity without local storage. This allocation pattern shows that ES siting is driven by system-wide network and temporal operating conditions rather than local DC capacity. Since Stage I-B minimizes only the total ES count, the table represents one minimum-unit siting solution.

\subsubsection{Frequency Security Status}

The maximum utilization ratios of the RoCoF, frequency-peak, and QSS limits are 0.208, 0.993, and 1.000, respectively. The QSS constraint binds, the frequency-peak constraint is nearly active, and RoCoF retains substantial margin.
The baseline DCLA is therefore limited by sustained frequency response capability, while the frequency-peak constraint is nearly active. An assessment based only on RoCoF would overestimate the available frequency security margin.

\subsubsection{Temporal Workload Scheduling}

\begin{figure}[!t]
\centering
\includegraphics[width=0.46\textwidth]{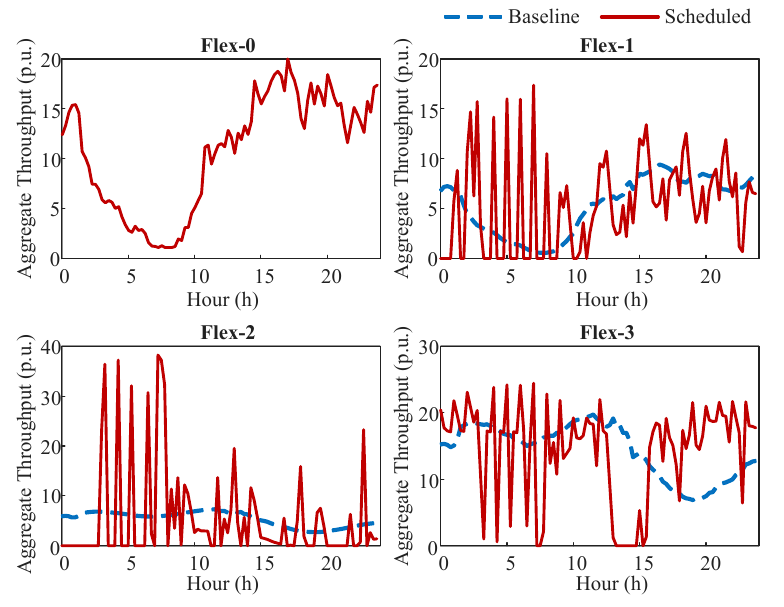}
\caption{Reference and optimized aggregate throughput for Flex-0--Flex-3. The dashed curves denote the reference execution profiles and the solid curves denote the QoS- and FS-aware schedules. Flex-0 remains rigid, whereas Flex-1--Flex-3 show progressively wider temporal shifting.}
\label{fig:flex_profiles}
\end{figure}

Fig.~\ref{fig:flex_profiles} compares the reference and optimized throughput profiles. Flex-0 follows its reference execution profile because it permits no deferral. Flex-1, Flex-2, and Flex-3 exhibit progressively wider temporal shifts as their service windows and QoS tolerances increase.
All schedules satisfy the rolling-window and daily completion constraints. The results confirm that longer service windows provide usable temporal flexibility, while the sharp interval-to-interval changes reflect the absence of an explicit throughput ramping constraint.

\subsection{Evaluation of Load Accommodation and Security Constraints}

This subsection compares four settings with different QoS and FS constraints to quantify the Grid-Centric DCLA bias and explain the resulting accommodation changes.

\subsubsection{Systematic Bias of Grid-Centric DCLA}

Table~\ref{tab:dcla_comparison} compares DCLA across four settings to quantify the systematic bias of Grid-Centric DCLA assessments. For the balanced mix, the Grid-Centric DCLA benchmark overestimates QoS- and FS-aware DCLA by 40.3\%. The main reduction comes from enforcing strict QoS, because rolling-window continuity and full daily completion prevent the benchmark from relying on implicit service curtailment or excessive intraday reshaping. FS alone is not binding under the relaxed service benchmark, since workloads can be shifted away from critical checkpoint periods. When strict QoS and FS are imposed together, this avoidance capability is reduced and the FS boundary further limits the QoS-feasible DCLA. This result shows that service feasibility and frequency security are coupled rather than independent constraints.

\begin{table}[!t]
\caption{DCLA by Constraint Setting and Workload Mix}
\label{tab:dcla_comparison}
\centering
\footnotesize
\renewcommand{\arraystretch}{1.2}
\setlength{\tabcolsep}{4pt}
\begin{tabular}{@{}lcccc@{}}
\hline\hline
\multirow{2}{*}{\textbf{Case}} & \multicolumn{3}{c}{\textbf{DCLA Across Workload Mixes (MW)}} & \multirow{2}{*}{\textbf{\begin{tabular}[c]{@{}c@{}}ES Units\\ (Balanced)\end{tabular}}} \\ \cline{2-4}
& \textbf{Inference-Heavy} & \textbf{Balanced} & \textbf{Training-Heavy} & \\
\hline
Grid-Centric & 630 & 720 & 750 & 16 \\
QoS-only     & 410 & 480 & 500 & 18 \\
FS-only     & 630 & 720 & 650 & 16 \\
Proposed     & 410 & 430 & 280 & 18 \\
\hline\hline
\end{tabular}
\par\vspace{2pt}
\parbox{\linewidth}{\scriptsize\textit{Note:} DCLA values are in MW. The ES column reports the balanced mix only. Grid-Centric uses relaxed daily adequacy; QoS-only enforces service constraints; FS-only enforces frequency-security constraints with relaxed service; and Proposed enforces both strict QoS and FS.}
\end{table}

\subsubsection{Frequency Security Effects of FS Constraints}

\begin{figure}[!t]
\centering
\includegraphics[width=1\columnwidth]{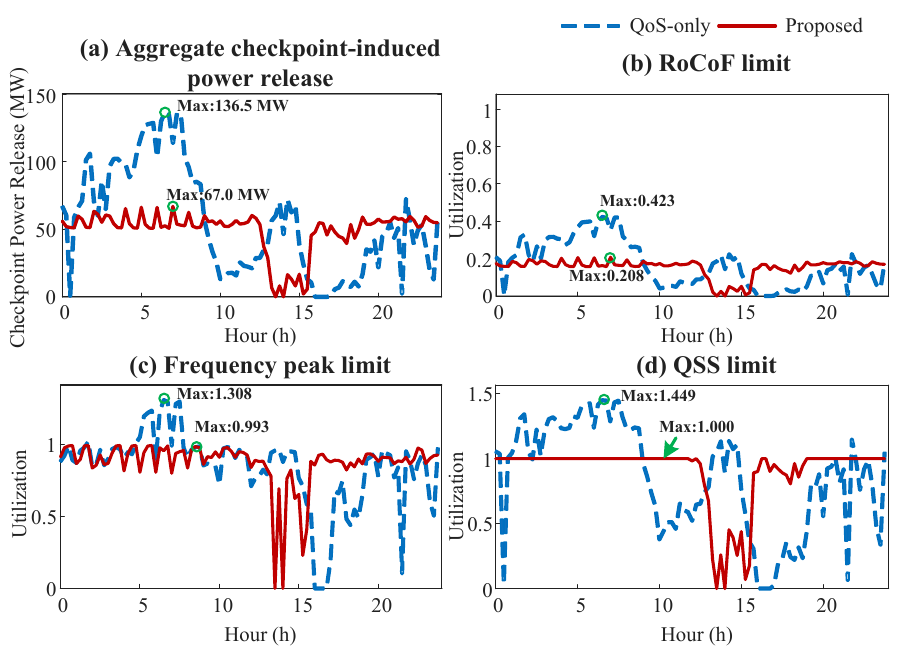}
\caption{Effect of FS constraints on checkpoint disturbances and frequency-security limits. Dashed curves show the QoS-only solution and solid curves show the proposed solution: (a) aggregate checkpoint-induced power release; (b) RoCoF-limit utilization; (c) frequency-peak-limit utilization; and (d) QSS-limit utilization. A utilization above 1 indicates a violation, and 1 denotes a binding limit.}
\label{fig:checkpoint_comparison}
\end{figure}

The impact of FS constraints on DCLA is examined by comparing the QoS-only and proposed solutions. Fig.~\ref{fig:checkpoint_comparison}(a) shows that the FS-aware formulation substantially suppresses the maximum checkpoint-induced power release. This reduction is much larger than the decrease in DCLA, indicating that the optimizer mainly improves frequency security by rescheduling checkpoint-enabled workloads and reducing synchronization, not merely by lowering accommodated capacity. Fig.~\ref{fig:checkpoint_comparison}(b)--(d) further shows that the QoS-only solution violates the frequency-peak and QSS limits, even though RoCoF remains below its limit. After imposing the FS boundary, the QSS constraint becomes binding and the frequency-peak constraint is nearly active. Therefore, the binding mechanism is associated with fast and sustained response adequacy rather than the initial inertial response alone.

\subsection{Sensitivity Analysis on Workload Characteristics}

Table~\ref{tab:dcla_comparison} also compares the inference-heavy, balanced, and training-heavy mixes under identical system settings.

\subsubsection{QoS-Limited DCLA Across Workload Mixes}

Without FS constraints, DCLA increases with the training-workload share. Greater deferrability improves service-feasible accommodation because training workloads can be shifted over longer windows and can better absorb network and resource constraints. Inference-dominated systems remain more constrained because a larger share of their demand is rigid or limited to short deferral windows.

\subsubsection{Interaction Between Workload Mix and FS}

The complete formulation changes this ranking. Training-heavy workloads provide the greatest temporal flexibility, but they also create the largest checkpoint-induced disturbance. As a result, the frequency-security penalty grows faster than the accommodation benefit from deferrability. Inference-heavy workloads cause weaker checkpoint disturbances, but their rigid service requirements limit the amount of usable flexibility. Under the studied settings, the balanced mix achieves the highest QoS- and FS-aware DCLA because it best balances service flexibility against frequency security risk.

\subsection{Complementary Effects of Workload Flexibility and Installable ES}

To evaluate the individual contributions of these two strategies, we conducted an ablation study by removing workload flexibility and installable ES independently while retaining all other settings.

\subsubsection{Contribution of Workload Flexibility}

When both workload flexibility and ES are disabled, the DCLA is limited to 330~MW. Introducing installable ES under rigid workloads yields no improvement, as the DCLA remains at 330~MW (with zero ES units being deployed). In that case, the active limitation comes from service inflexibility rather than from transferable grid or energy margins. Workload flexibility is therefore the primary enabler of additional accommodation, increasing the DCLA to 400~MW even without the use of ES. By allowing throughput to move within service windows, flexible scheduling creates feasible operating space that cannot be obtained by storage alone.

\subsubsection{Additional Contribution of Installable ES}

With flexible scheduling enabled, installable ES provides an additional accommodation gain, bringing the maximum DCLA to 430~MW through the deployment of 18 ES units. In the studied system, storage complements workload flexibility rather than replacing it. Flexible scheduling first creates temporal accommodation headroom, after which ES exploits the remaining system-wide network and intertemporal margins.

\section{Conclusion}  

This paper developed an optimization-based DCLA method that jointly accounts for grid adequacy, workload QoS, and checkpoint-induced FS. A QoS-aware workload model was developed to characterize heterogeneous workload flexibility while ensuring service continuity and workload completion. Checkpoint-induced power releases were linked to accommodated training workloads, and the corresponding RoCoF, frequency-peak, and QSS limits were reformulated as algebraic constraints. These components were then integrated into a two-stage lexicographic formulation, in which Stage I maximizes DCLA and minimizes the required ES deployment, while Stage II optimizes system operation with the DC and ES siting and sizing decisions fixed.

Case studies on the IEEE 39-bus system show that conventional grid-adequacy-based DCLA methods overestimate QoS- and FS-aware DCLA by 40.3\% for the balanced workload mix. QoS accounts for the largest capacity reduction by limiting excessive workload reshaping and preventing implicit workload curtailment, whereas checkpoint-induced FS constraints become binding for training-heavy workloads. In the baseline case, the QSS limit is binding and the frequency-peak limit is nearly active, while RoCoF retains substantial margin.

The results reveal the dual effect of workload flexibility on DCLA. Greater deferrability can alleviate network and intertemporal constraints, but the associated checkpoint-induced power releases may offset these benefits and make frequency security the binding constraint. Therefore, greater workload flexibility does not necessarily increase secure DCLA. Reliable DC interconnection decisions should jointly account for the grid-support benefits of workload flexibility, its QoS limitations, and the associated frequency-security risks. Future work will incorporate more detailed workload and checkpoint dynamics, operational uncertainty, and higher-fidelity frequency-response models.

\end{document}